\documentstyle[12pt]{article}
\newenvironment{th}[1]{\trivlist \item[\hskip \labelsep{\bf Theorem #1.}]}
{\endtrivlist}

\newenvironment{cor}{\trivlist \item[\hskip \labelsep{\bf Corollary.}]\it}
{\endtrivlist}
\newenvironment{dfn}[1]{\trivlist \item[\hskip \labelsep{\bf Definition #1.}]}
{\endtrivlist}
{\endtrivlist}
{\endtrivlist}
\begin{document}
\title{\bf Nambu structures on four dimensional real Lie groups and related superintegrable systems}
\author{S. Farhang-Sardroodi$^a$\thanks{e-mail:m.farhang88@azaruniv.edu},\,\, A. Rezaei-Aghdam$^{a}$\thanks{e-mail:rezaei-a@azaruniv.edu}\,\,and L. Sedghi-Ghadim$^{b}$\thanks{e-mail:m.sedghi88@azaruniv.edu}\\ \\
{\small {$^a${\em Department of Physics, Azarbaijan Shahid Madani
University , 51745-406, Tabriz, Iran.}}}\\
{\small {$^b${\em Department of Mathematics, Azarbaijan Shahid Madani
University , 51745-406, Tabriz, Iran.}}}}

\maketitle
\begin{abstract}
We have determined  all Nambu tensors (Nambu structures) of order
four and three on four dimensional real Lie groups. Furthermore, we have
obtained superintegrable systems by use of the Nambu structures of order
four on  some of these Lie groups as phase spaces with symmetry groups ${\bf A_{4,8}}$ and ${\bf A_{4,10}}$.
\end{abstract}
\section{\bf Introduction}
In 1973 \cite{Nam} Nambu studied a dynamical system which was
defined as a Hamiltonian system with respect to a generalization of
Poisson bracket (Poisson-like bracket), defined by a Jacobian
determinant. Some years (about two decades) later Takhtajan
\cite{Tak} introduced the concept of Nambu-Poisson (or simply Nambu)
structure using an axiomatic formulation for $n$-bracket and gave
the basic properties of this operation and also geometric
formulations of Nambu manifolds. This new approach motivated a
series of papers about some new concepts.
Nambu manifold is a {\small$C^{\infty}$} manifold endowed with Nambu
tensor, a skew-symmetric contravariant tensor field on a manifold
such that the induced bracket operation satisfies the fundamental
identity, which is a generalization of the usual Jacobi
identity \cite{Gau}-\cite{Nak1} (there are another
generalization so-called generalized Poisson bracket
\cite{Azc1,Azc2}; where a comparison of both concepts was given in
\cite{Azc3,Ib1}). In \cite{Gra2} and \cite{Vai} the
concept of Nambu Lie group was presented. In \cite{Vai} Vaisman
extended Nambu brackets to 1-forms and by generalizing the
Poisson-Lie case, he defined Nambu-Lie groups as the Lie groups
which were endowed with a multiplicative Nambu structure. The
decomposability of Nambu structures for Lie groups and also
the correspondence between the set of left invariant Nambu tensors of
order $n$ on $m$ dimensional Lie groups {\small$G$} with the set of $n$
dimensional Lie subalgebras of {\small${\cal{G}}$} (Lie algebra of {\small$G$})
were proven in \cite{Nak2} by Nakanishi. He also determined the
multiplicative Nambu structures on three dimensional real Lie groups
in  \cite{Nak3}. In this paper as the same way we will try to determine the multiplicative Nambu
structure of order four and three on four dimensional real Lie
groups. The outline of the paper is as follows:
\\In section two, for selfcontianing of the paper we review some
definitions and theorems. Then, in section three, by means of the
method applied in  \cite{Nak3} we determine the multiplicative Nambu
strutures of order four and three on the real four dimensional Lie
groups. Finally, in section four, by use of the Nambu structures of
order four on  some Lie groups (i.e. four dimensional real Lie groups that have symplectic structures
\cite{O}) we obtain superintegrable systems with these Lie groups as phase spaces and ${\bf A_{4,8}}$
and ${\bf A_{4,10}}$ as symmetry Lie groups.
\section{\bf Basic definitions and theorems}
For self containing of the paper let us recall some basic
definitions and theorems about Nambu structure (\cite{Gra2,Nak3}).\\
 Let {\small$G$} be an $m$ dimensional Lie group with Lie algebra
{\small${\mathcal{G}}$}. Denote {\small$\Gamma(\Lambda^nTG)$} as the set of
antisymmetric $n$-vector fields (contravariant tensors) on {\small$G$}. Then
for each {\small$\eta\in \Gamma(\Lambda^nTG)$} one can define an $n$-bracket
of the functions on Lie group {\small$G$} as follows:
{\small
\begin{eqnarray}
\{f_1,...,f_n\}=\eta(df_1,...,df_n),\quad\forall f_i \in {\cal F}(G),\quad i=1,\cdots,n,
\end{eqnarray}
}
where {\small${\cal{F}}(G)$} is the algebra of $c^{\infty}$ functions on {\small$G$}.\\
Furthermore, since the bracket satisfies Leibniz rule, one can
define a vector field {\small$X_{f_1,...,f_{n-1}}$}by
{\small
\begin{eqnarray}
X_{f_1,...,f_{n-1}}(\textit{g})=\{f_1,...,f_{n-1},\textit{g}\},\quad\forall{\textit{g}}\in
{\cal F}(G),
\end{eqnarray}
}
where this vector field is called \emph{Hamiltonian }vector field; the space of \emph{Hamiltonian} vector fields is denoted by {\small$\cal{H}$}.
\begin{dfn}{1} \cite{Nak1,Nak2,Nak3} \emph{{An
element {\small$\eta\in \Gamma(\Lambda^nTG)$}, for $n\geq 3$ is called a
Nambu tensor of order n if it satisfies {\small${\cal L}_{X_{f_1,...,f_{n-1}}}\eta=0$, $\forall X_{f_{1},...,f_{n-1}}$ $\in \cal{H}$}  with {\small$f_i\in{\cal F}(G)$}; where {\small${\cal L}$} stands for Lie
derivative.}}
\end{dfn}
\begin{dfn}{2} \cite{Nak1,Nak2,Nak3} \emph{ An
element {\small$\eta\in \Gamma(\Lambda^nTG)$} is
said to be a multiplicative tensor if {\small$\forall g_1,g_2\in G$}, we have
{\small
\begin{eqnarray}
\eta_{g_1g_2}={L_{g_1}}_\ast\eta_{g_2}+{R_{g_2}}_\ast\eta_{g_1},
\end{eqnarray}
}
where {\small$R_{g_2}$} and {\small$L_{g_1}$} are right and left translations in
{\small$G$} respectively.\\} A Lie group {\small$G$} endowed with  a
multiplicative Nambu tensor $\eta$ is
called \emph{Nambu-Lie group}\cite{Vai}.
\end{dfn}
\begin{th}{1} \cite{Vai} \emph{Let {\small$G$} be an $m$-dimensional Lie
group, and Let $\textbf{h}$ be an $n$-dimensional Lie subalgebra of}
{\small${\mathcal{G}}$} \emph{with $n\geq 3$, for a basis {\small$\{X_1,...,X_n\}$} of
$\textbf{h}$ , put {\small$\eta = X_1\wedge...\wedge X_n$}. Then $\eta$ is
left invariant Nambu tensor of order $n$ on {\small$G$}. Conversely, given a
left invariant Nambu tensor }{\small$\eta = X_1\wedge...\wedge X_n \in
\Lambda^n {\mathcal{G}}$} \emph{on {\small$G$}, then
{\small$\textbf{h}=\{X_1,...,X_n\}$} is a Lie subalgebra of} {\small${\mathcal{G}}$}.
\end{th}
\begin{cor} \cite{Vai} There is a one to one
correspondence up to a coefficient (function) between the set of left
invariant Nambu tensors of order $n$ on {\small$G$} and the set of
$n$-dimensional Lie subalgebras of {\small${\mathcal{G}}$}.
\end{cor}
Notice that for a Nambu tensor $\eta$ of order $n\geq 3$, if $f$ is a
smooth function, then $f\eta$ is again a Nambu
tensor \cite{Nak1}.
\begin{th}{2} \cite{Gra1}\emph{ Let {\small$(G,\eta)$} be
an $m$-dimensional compact or semisimple Nambu-Lie group, and let
$\eta$ be of top order, then $\eta=0$.\\} The following theorem
gives one of the characterizations of
Nambu-Lie groups, which was proved by Vaisman \cite{Vai}.
\end{th}
\begin{th}{3} \cite{Vai} \emph{If {\small$G$} is a connected Lie group
endowed with a Nambu tensor $\eta$ which vanishes at the unit $e$
of {\small$G$}, then {\small($G,\eta$)} is a Nambu-Lie group if and only if the
$n$-bracket of any $n$ left (right) invariant 1-forms of {\small$G$} is a
left (right) invariant 1-form.}
\end{th}
 Using the above theorem one
can characterize a multiplicative tensor $\eta$ of top order. Let
{\small${\mathcal{G}}$} be a Lie algebra of {\small$G$} with
 basis {\small$X_1,\cdots , X_m$}. It is clear that the left invariant vector
 fields can be considered as basis for {\small${\mathcal{G}}$}, we denote these left invariant vector fields
 by the same letters $X_i$.
 Since $\eta$ is of top order, $\eta$ has an expression
 {\small$\eta=f X_1\wedge \cdots \wedge X_m\quad \mbox{for some  f}\in{ \cal F}(G)$}.\\
 According to these notations we have:
 \begin{th}{4} \cite{Nak3} \emph{Let {\small$\eta=fX_1\wedge...\wedge X_m$,\quad $f\in{\cal F}(G)$}
 be a tensor of top order on {\small$G$} (such a tensor is always a Nambu tensor).
Then $\eta$ is multiplicative if and only if {\small$f(e)=0\;$}and
{\small
\begin{eqnarray}
X_if+(\sum_{k=1}^mC_{ik}^k)f=q_i\;\;\;i=1,...,m,
\end{eqnarray}
}
where {\small${C_{ij}^k}$} is structure constant of {\small${\mathcal{G}}$}
with respect to the basis {\small$X_1,...,X_m$}, and $q_i (i=1,...,m)$ are
some constants.}
\end{th}
In \cite{Nak3} using the above theorem the
Nambu structures of  order three for the three dimensional real Lie
groups were obtained. In this paper, in the same way we have determined Nambu
structure of order four (top order) for four dimensional real Lie
groups; and also by the use of the theorem 1, we  have calculated Nambu structure
of order three for these Lie groups.
\section{\bf Nambu structures on four dimensional real Lie groups}
In this section, by means of the theorems 1 and 4, we calculate Nambu
structures of order four and three on four dimensional real Lie groups.
 Note that we use the Patera and  Winternitz classification \cite{Pat}
 for four dimensional Lie algebras and their subalgebras.\\
We denote by {\small${\mathcal{G}}$} the four-dimensional real Lie
algebra, corresponding to the simply connected Lie group {\small$G$}; and also
the left invariant vector fields are denoted by
{\small$X_1,X_2,X_3,X_4$}. To calculate these vector fields, we need to determine the
left invariant 1-forms, where already in \cite{Rez} these calculations were
performed. Here we use those results for obtaining the left
invariant vector fields. In general, for a Lie group {\small$G$} with Lie algebra
{\small${\mathcal{G}}$} with abstract basis {\small$\{T_i\}$} the left invariant one forms can be determined as follows:
{\small
\begin{eqnarray}
g^{-1}dg=e^i\;_\mu T_i dx^\mu,\quad\forall g \in G,\nonumber
\end{eqnarray}
}
such that for the left invariant vector fields we have:
{\small
\begin{eqnarray}
X_i=V_i\;^\mu \partial_\mu,\nonumber
\end{eqnarray}
}
where {\small$V_i\;^\mu=(e_i\;^\mu)^{t}$}, and $e_i\;^\mu$ is inverse of $e^i\;_\mu$ and $t$ stands for
transpose of the matrix that has already been obtained in \cite{Rez}; so one can calculate the left invariant
vector fields. The results are written in table 1.\\
Now for calculating the Nambu structure {\small $\eta\in \Gamma(\Lambda^4TG)$} one can write it as
{\small \emph{$\eta=fX_1\wedge X_2\wedge X_3\wedge X_4$}}, and for {\small $\eta\in \Gamma(\Lambda^3TG)$}
it can be as {\small $\eta=fX_1\wedge X_2\wedge X_3$}, such that {\small $f\in {\mathcal{F}}(G)$}. Then
using the theorem 4, we calculate the Nambu structures of order
four on four dimensional real Lie groups and also by  use of the
theorems 1 and 4, we obtain the Nambu structures of order three on
real four dimensional Lie groups. Before listing the results, for presentation of the method,
let us apply this method on the Lie algebra $A_{4,8}$ as an example.
This is a Lie algebra which is isomorphic to Heisenberg algebra
and we have the following commutative relations for it \cite{Pat}:
{\small
\begin{eqnarray}
&&\hspace{-5mm}[X_2,X_4]=X_2,\ [X_3,X_4]=-X_3,\
[X_2,X_3]=X_1.
\end{eqnarray}
}
The left invariant vector fields for this Lie algebra are obtained as
{\small
\begin{eqnarray}
X_1=\partial_1,\quad
X_2=(-x^3e^{-x^4})\partial_1+e^{-x^4}\partial_2,\quad
X_3=e^{x^4}\partial_3,\quad
X_4=\partial_4,\nonumber
\end{eqnarray}
}
such that these vector fields satisfy the commutation relations (5). From the theorem 4, a function
{\small$f(x^1,x^2,x^3,x^4)$} must satisfy {\small$f(0,0,0,0)=0$} and
{\small
\begin{eqnarray}
X_if+(\sum_{k=1}^mC_{ik}^k)f=q_i\;\;\;i=1,...,4,\label{Nam-cal}
\end{eqnarray}
}
where $q_i$ are some constants. In this way one can obtain a solution of $(\ref{Nam-cal})$ as
{\small$f=q_4x^4$}, so that we have
{\small
\begin{eqnarray}
&&\hspace{-5mm}\eta=q_4x^4\partial_1\wedge
\partial_2\wedge \partial_3\wedge \partial_4,\nonumber
\end{eqnarray}
}
which gives a Nambu-Lie  structure of order four on the
corresponding Lie
group {\small$\bf{A_{4,8}}$}.\\
In the same way, for three dimensional Lie subalgebras of $A_{4,8}$ \cite{Pat} we have the following left invariant vector fields:
{\small
\begin{eqnarray}
&& A_{3,1}:\{X_2,X_3;X_1\},\nonumber\\
&& X_2=\partial_2-x_3\partial_1,\quad\quad
X_3=\partial_3,\quad\quad
X_1=\partial_1,\nonumber
\end{eqnarray}
\begin{eqnarray}
&& A_2\oplus A_1:\{X_4,X_1;X_2\},\nonumber\\
&&X_4=\partial_4+x_2\partial_2,\quad\quad
X_1=\partial_1,\quad\quad
X_2=\partial_2,\nonumber
\end{eqnarray}
\begin{eqnarray}
&& A_2\oplus A_1:\{X_4,X_1;X_3\},\nonumber\\
&& X_4=\partial_4-x_3\partial_3,\quad\quad
X_1=\partial_1,\quad\quad
X_3=\partial_3,\nonumber
\end{eqnarray}
}
so we have the following Nambu structures of order three on {\small$\bf{A_{4,8}}$} (respectively):
{\small
\begin{eqnarray}
&&\hspace{-5mm}\eta_1=(q_1x^2+q_2x^3)\partial_1\wedge \partial_2\wedge
\partial_3,\nonumber\\
&&\hspace{-5mm}\eta_2=(q_3x^2+q_1(e^{x^4}-1))\partial_1\wedge \partial_2\wedge
\partial_4,\nonumber\\
&&\hspace{-5mm}\eta_3=(q_3x^3+q_1(e^{-x^4}-1))\partial_1\wedge \partial_3\wedge
\partial_4,\nonumber
\end{eqnarray}
}
In this way, we have determined all Nambu structures of order four and
three on four dimensional real Lie groups. The results have been listed in table I and II :\\
 \hspace{-0.5cm}{\footnotesize  TABLE I.} {\footnotesize
Nambu structures of order four on four dimensional real Lie groups
and the corresponding Left invariant
vector fields.}\\
\begin{tabular}{l l l l  p{15mm} }
    \hline\hline
{\scriptsize Lie algebra }& {\scriptsize Structure
constants}&{\scriptsize Left invariant vector fields}
& {\scriptsize $\eta^{1234}$}  \smallskip\\
\hline

{\scriptsize $4A_1$}& {\scriptsize $f^k_{ij}=0$}&
{\scriptsize$X_1=\partial_1,\quad
X_2=\partial_2
$}& {\scriptsize$q_1x^1+q_2x^2+q_3x^3+q_4x^4$} \\

{\scriptsize $$}&{\scriptsize $ $}&
{\scriptsize$X_3=\partial_3,\quad X_4=\partial_4$}& {\scriptsize$$}\\
\hline

{\scriptsize $A_2\oplus2A_1$}&{\scriptsize $f^2_{12}=1 $}&
{\scriptsize$ X_1=\partial_1-x^2\partial_2,\quad
X_2=\partial_2 $}& {\scriptsize$q_2x^2+q_1(e^{-x^1}-1)$}\\

{\scriptsize $$}&{\scriptsize $ $}&
{\scriptsize$X_3=\partial_3,\quad X_4=\partial_4 $}& {\scriptsize$$}\\
\hline

{\scriptsize $2A_2$}&{\scriptsize $f^2_{12}=1,f^4_{34}=1$}&
{\scriptsize$ X_1=\partial_1-x^2\partial_2,\quad X_2=\partial_2 $}& {\scriptsize $q_3(e^{-(x^1+x^3)}-1)$}\\

{\scriptsize $$}&{\scriptsize $$}& {\scriptsize
$X_3=\partial_3-x^4\partial_4,\quad X_4=\partial_4$}& {\scriptsize$$}\\
\hline

{\scriptsize $A_{3,1}\oplus A_1$}&{\scriptsize $f^1_{23}=1$}&
{\scriptsize $X_1=\partial_1,\quad
X_2=-x^3\partial_1+\partial_2 $}& {\scriptsize$q_2x^2+q_3x^3+q_4x^4$}\\
\hline\hline
\end{tabular}
\vspace{8mm}
\newpage
\hspace{-0.5cm}{\footnotesize  TABLE I.}  {\footnotesize
Nambu structures of order four on four dimensional real Lie groups
and  the corresponding Left invariant
vector fields (continue).}\\
    \begin{tabular}{l l l l  p{15mm} }
    \hline\hline
{\scriptsize Lie algebra }& {\scriptsize Structure
constants}&{\scriptsize Left invariant vector fields}
& {\scriptsize $\eta^{1234}$}  \smallskip\\
\hline
\smallskip

{\scriptsize $A_{4,1}$}&{\scriptsize $f^1_{24}=1,f^2_{34}=1 $}&
{\scriptsize$X_1=\partial_1,\quad
X_2=-x^4\partial_1+\partial_2$}& {\scriptsize$q_3x^3+q_4x^4$} \\

{\scriptsize $$}&{\scriptsize $$}&
{\scriptsize$X_3=\frac{1}{2}(x^4)^2\partial_1-x^4\partial_2+\partial_3$}
{\scriptsize $$}\\

{\scriptsize $$}&{\scriptsize $$}&
{\scriptsize$X_4=\partial_4$}
{\scriptsize $$}\\
\hline
{\scriptsize $A^a_{4,2}$}&{\scriptsize
$f^1_{14}=a,f^2_{24}=1$} &
{\scriptsize$X_1=e^{-ax^4}\partial_1,\quad
X_2=e^{-x^4}\partial_2$}
& {\scriptsize$q_4(e^{-(a+2)x^4}-1)$} \\

{\scriptsize $$}&{\scriptsize $f^2_{34}=1,f^3_{34}=1$}&
{\scriptsize$X_3=-x^4e^{-x^4}\partial_2+e^{-x^4}\partial_3$}
{\scriptsize $$}\\

{\scriptsize $$}&{\scriptsize $$}&
{\scriptsize$X_4=\partial_4$}
{\scriptsize $$}\\
\hline
{\scriptsize $A^1_{4,2}$}&{\scriptsize
$f^1_{14}=1,f^2_{24}=1 $}&
{\scriptsize$X_1=e^{-x^4}\partial_1,\quad
X_2=e^{-x^4}\partial_2$}& {\scriptsize$q_4(e^{-3x^4}-1)$} \\
\hline \hline
\end{tabular}
\newpage
\vspace{8mm}
\hspace{-0.5cm}{\footnotesize  TABLE I.} {\footnotesize
Nambu structures of order four on four dimensional real Lie groups
and  the corresponding Left invariant
vector fields (continue).}\\
    \begin{tabular}{l l l l  p{15mm} }
    \hline\hline
{\scriptsize Lie algebra }& {\scriptsize Structure
constants}&{\scriptsize Left invariant vector fields}
& {\scriptsize $\eta^{1234}$}  \smallskip\\
\hline
\smallskip

{\scriptsize $A_{4,7}$}&{\scriptsize
$f^1_{14}=2,f^2_{24}=1,f^1_{23}=1$}&
{\scriptsize$X_1=e^{-2x^4}\partial_1
$}& {\scriptsize$q_4(e^{-4x^4}-1)$} \\

{\scriptsize $$}&{\scriptsize $f^2_{34}=1,f^3_{34}=1$}&
{\scriptsize$X_2=x^3e^{-x^4}\partial_1$}
{\scriptsize $$}\\

{\scriptsize $$}&{\scriptsize $$}& {\scriptsize$+e^{-x^4}\partial_2$}
{\scriptsize $$}\\

{\scriptsize $$}&{\scriptsize $$}&
{\scriptsize$X_3=-x^3x^4e^{-x^4}\partial_1 $}
{\scriptsize $$}\\

{\scriptsize $$}&{\scriptsize $$}&
{\scriptsize$-x^4e^{-x^4}\partial_2+e^{-x^4}\partial_3$}
{\scriptsize $$}\\

{\scriptsize $$}&{\scriptsize $$}&
{\scriptsize$X_4=\partial_4$}
{\scriptsize $$}\\
\hline

{\scriptsize $A_{4,8}$}&{\scriptsize
$f^1_{23}=1,f^2_{24}=1,f^3_{34}=-1 $}&
{\scriptsize$X_=\partial_1
$}& {\scriptsize$q_4x^4$} \\
\hline\hline
\end{tabular}
\newpage
\vspace{9mm}
 \hspace{-0.5cm}{\footnotesize TABLE II.} \footnotesize
{Corresponding Lie subalgebras with left invariant vector fields
and Nambu structures of order three on Lie groups.}\\
    \begin{tabular}{l l l  l p{15mm} }
    \hline\hline
{\scriptsize Lie subalgebra }& {\scriptsize Lie subalgebra
basis}&{\scriptsize Left invariant vector fields}& {\scriptsize
$\eta^{ijk}$}  \smallskip\\
\hline
\smallskip

\vspace{-2mm}

{\scriptsize $3A_1\subset 4A_1$}& {\scriptsize $X_1+aX_4,X_2+bX_4$}&
{\scriptsize $X_1+aX_4=\partial_1+a\partial_4$}& {\scriptsize
$\eta^{123}=q_1x^1+q_2x^2+q_3x^3$}\vspace{2mm}\\
{\scriptsize $$}&{\scriptsize $X_3+cX_4$} &{\scriptsize
$X_2+bX_4=\partial_2+b\partial_4$}&{\scriptsize $\eta^{124}=c(q_1x^1+q_2x^2+q_3x^3)$}\vspace{2mm}\\

{\scriptsize $$}&  {\scriptsize $$}& {\scriptsize
$X_3+cX_4=\partial_3+c\partial_4$}& {\scriptsize $\eta^{143}=b(q_1x^1+q_2x^2+q_3x^3)$}\\

{\scriptsize $$}& {\scriptsize $$}& {\scriptsize $
$}& {\scriptsize $\eta^{423}=a(q_1x^1+q_2x^2+q_3x^3)$} \\

{\scriptsize $$}& {\scriptsize $$}& {\scriptsize $$}& {\scriptsize $$} \\

{\scriptsize $$}&{\scriptsize $X_1+aX_2,X_3,X_4$}&{\scriptsize
$X_1+aX_4=\partial_1+a\partial_4$}
&{\scriptsize $\eta^{134}=q_1x^1+q_2x^3+q_3x^4$} \\

{\scriptsize $$}&{\scriptsize $$}&{\scriptsize
$X_3=\partial_3,\quad
X_4=\partial_4$}&{\scriptsize $\eta^{234}=a(q_1x^1+q_2x^3+q_3x^4)$} \\
\hline\hline
\end{tabular}
\newpage
\vspace{9mm}
 \hspace{-0.5cm}{\footnotesize TABLE II.} {\footnotesize
{ Corresponding Lie subalgebras with left invariant vector fields
and Nambu structures of order three on Lie groups (continue).}}\\
    \begin{tabular}{l l l  l p{15mm} }
    \hline\hline
{\scriptsize Lie subalgebra }& {\scriptsize Lie subalgebra
basis}&{\scriptsize Left invariant vector fields}& {\scriptsize
$\eta^{ijk}$}  \smallskip\\
\hline
\smallskip

\vspace{-2mm}


{\scriptsize $3A_1\subset A_{3,1}\oplus A_1$}&{\scriptsize
$X_1,X_2\cos\varphi$}&{\scriptsize $X_1=\partial_1$}& {\scriptsize $\eta^{124}
=(q_1x^1+q_2((x^3)^2$}\\

{\scriptsize $$}&{\scriptsize $+X_3\sin\varphi,X_4$}&{\scriptsize
$X_2\cos\varphi+X_3\sin\varphi=\cos\varphi\partial_2+\sin\varphi\partial_3
 $}&{\scriptsize $+(x^2)^2)^{\frac{1}{2}}+q_3x^4)\cos\varphi$} \\

{\scriptsize $$}&{\scriptsize $$}&{\scriptsize
$X_4=\partial_4$}&{\scriptsize$ \eta^{134}=q_1x^1+q_2((x^3)^2)$} \\

{\scriptsize $$}&{\scriptsize $$}&{\scriptsize
$$}&{\scriptsize $+(x^2)^2)^{\frac{1}{2}}+q_3x^4)\sin\varphi$} \\

{\scriptsize $$}& {\scriptsize $$}& {\scriptsize $$}& {\scriptsize $$} \\

{\scriptsize $A_{3,1}\subset A_{3,1}\oplus A_1$}&{\scriptsize
$X_2+aX_4,X_3+bX_4;X_1$}&{\scriptsize
$X_2+aX_4=\partial_2+a\partial_4-x^3\partial_1
$}&{\scriptsize $\eta^{231} =q_1x^2+q_2x^3$} \\

{\scriptsize $$}&{\scriptsize $$}&{\scriptsize
$X_3+bX_4=\partial_3+b\partial_4,\quad X_1=\partial_1$}&{\scriptsize $\eta^{231} =b(q_1x^2+q_2x^3)$} \\
\hline\hline
\end{tabular}

\newpage
\vspace{9mm}

 \hspace{-0.5cm}{\footnotesize TABLE II.} {\footnotesize
{ Corresponding Lie subalgebras with left invariant vector fields
and Nambu structures of order three on Lie groups (continue).}}\\
    \begin{tabular}{l l l  l p{15mm} }
    \hline\hline
{\scriptsize Lie subalgebra }& {\scriptsize Lie subalgebra
basis}&{\scriptsize Left invariant vector fields}& {\scriptsize
$\eta^{ijk}$}  \smallskip\\
\hline
\smallskip

\vspace{-2mm}


{\scriptsize $3A_1\subset A_{3,6}\oplus A_1$}&{\scriptsize
$X_1,X_2,X_4$}&{\scriptsize $X_1=\partial_1,\quad
X_2=\partial_2,\quad
X_4=\partial_4$}&{\scriptsize $\eta^{124}
=q_1x^1+q_2x^2+q_3x^4$} \\

{\scriptsize $$}&{\scriptsize $$}&{\scriptsize $$}&{\scriptsize $$} \\

{\scriptsize $A_{3,6}\subset A_{3,6}\oplus A_1$}&{\scriptsize
$X_3+aX_4;X_1,X_2$}&{\scriptsize $X_3+aX_4=\partial_3+a\partial_4$}&{\scriptsize $\eta^{123}
=\frac{1}{2}q_1(x^3+\frac{1}{a}x^4)$}\\

{\scriptsize $$}&{\scriptsize $$}&{\scriptsize
$+x^2\partial_1-x^1\partial_2$}&{\scriptsize $\eta^{124} =\frac{1}{2}q_1(x^3+\frac{1}{a}x^4)$} \\

{\scriptsize $$}&{\scriptsize $$}&{\scriptsize
$X_1=\partial_1,\quad X_2=\partial_2$}&{\scriptsize $$} \\
\hline

{\scriptsize $3A_1\subset A^a_{3,7}\oplus A_1$}&{\scriptsize
$X_1,X_2,X_4$}&{\scriptsize $X_1=\partial_1,\quad
X_2=\partial_2,\quad X_4=\partial_4$}&{\scriptsize $\eta^{124}=q_1x^1+q_2x^2+q_3x^4$} \\

{\scriptsize $$}&{\scriptsize $$}&{\scriptsize $$}&{\scriptsize $$} \\
\hline\hline
\end{tabular}

\newpage
\vspace{9mm}

 \hspace{-0.5cm}{\footnotesize TABLE II.} {\footnotesize
{ Corresponding Lie subalgebras with left invariant vector fields
and Nambu structures of order three on Lie groups (continue).}}\\
    \begin{tabular}{l l l  l p{15mm} }
    \hline\hline
{\scriptsize Lie subalgebra }& {\scriptsize Lie subalgebra
basis}&{\scriptsize Left invariant vector fields}& {\scriptsize
$\eta^{ijk}$}  \smallskip\\
\hline
\smallskip

\vspace{-2mm}

{\scriptsize$3A_1\subset
A_{4,4}$}&{\scriptsize$X_1,X_2,X_3$}&{\scriptsize$X_1=\partial_1,\quad X_2=\partial_2,\quad X_3=\partial_3$}&{\scriptsize$\eta^{123}=q_1x^1+q_2x^2+q_3x^3$}\\

{\scriptsize $$}&{\scriptsize $$}&{\scriptsize $$}&{\scriptsize $$} \\

{\scriptsize$A_{3,2}\subset
A_{4,4}$}&{\scriptsize$X_4;X_1,X_2$}&{\scriptsize$X_4=\partial_4+(x^1+x^2)\partial_1+x^2\partial_2
$}&{\scriptsize$\eta^{124}=q_1(e^{2x^4}-1)$}\\

{\scriptsize$$}&{\scriptsize$$}&{\scriptsize$X_1=\partial_1,\quad X_2=\partial_2$}&{\scriptsize$$}\\
\hline

{\scriptsize$3A_1\subset
A^{a,b}_{4,5}$}&{\scriptsize$X_1,X_2,X_3$}&{\scriptsize$X_1=\partial_1,\quad X_2=\partial_2,\quad X_3=\partial_3$}&{\scriptsize$\eta^{123}=q_1x^1+q_2x_2+q_3x^3$}\\

{\scriptsize $$}&{\scriptsize $$}&{\scriptsize $$}&{\scriptsize $$} \\

{\scriptsize$A^a_{3,5}\subset
A^{a,b}_{4,5}$}&{\scriptsize$X_4;X_1,X_2$}&{\scriptsize$X_4=\partial_4+x^1\partial_1+(ax^2)\partial_2$}&{\scriptsize$\eta^{124}=q_1(e^{(a+1)x^4}-1)$}\\
\hline\hline
\end{tabular}
\newpage
\vspace{9mm}
 \hspace{-0.5cm}{\footnotesize TABLE II.} {\footnotesize
{ Corresponding Lie subalgebras with left invariant vector fields
and Nambu structures of order three on Lie groups (continue).}}\\
    \begin{tabular}{l l l  l p{15mm} }
    \hline\hline
{\scriptsize Lie subalgebra }& {\scriptsize Lie subalgebra
basis}&{\scriptsize Left invariant vector fields}& {\scriptsize
$\eta^{ijk}$}  \smallskip\\
\hline
\smallskip

\vspace{-2mm}


{\scriptsize$A_{3,1}\subset
A_{4,7}$}&{\scriptsize$X_2,X_3;X_1$}&{\scriptsize$X_2=\partial_2-x^3\partial_1$}&{\scriptsize$\eta^{123}=q_1x^2+q_2x^3$}\\

{\scriptsize$$}&{\scriptsize$$}&{\scriptsize$
X_3=\partial_3
,\quad X_1=\partial_1$}&{\scriptsize$$}\\

{\scriptsize $$}&{\scriptsize $$}&{\scriptsize $$}&{\scriptsize $$} \\

{\scriptsize$A^{{1}/{2}}_{3,5}\subset
A_{4,7}$}&{\scriptsize$X_4;X_1,X_2$}&{\scriptsize$X_4=\partial_4+2x^1\partial_1+x^2\partial_2$}&{\scriptsize$\eta^{124}=q_1(e^{3x^4}-1)$}\\

{\scriptsize$$}&{\scriptsize$$}&{\scriptsize$
X_1=\partial_1,\quad X_2=\partial_2$}&{\scriptsize$$}\\
\hline
{\scriptsize$A_{3,1}\subset
A_{4,8}$}&{\scriptsize$X_2,X_3;X_1$}&{\scriptsize$X_2=\partial_2-x^3\partial_1$}&{\scriptsize$\eta^{123}=q_2x^2+q_2x^3$}\\

{\scriptsize$$}&{\scriptsize$$}&{\scriptsize$
X_3=\partial_3
,\quad X_1=\partial_1$}&{\scriptsize$$}\\
\hline\hline
\end{tabular}
\newpage
\vspace{9mm}
 \hspace{-0.5cm}{\footnotesize TABLE II.} {\footnotesize
{ Corresponding Lie subalgebras with left invariant vector fields
and Nambu structures of order three on Lie groups (continue).}}\\
    \begin{tabular}{l l l  l p{15mm} }
    \hline\hline
{\scriptsize Lie subalgebra }& {\scriptsize Lie subalgebra
basis}&{\scriptsize Left invariant vector fields}& {\scriptsize
$\eta^{ijk}$}  \smallskip\\
\hline
\smallskip

\vspace{-2mm}
{\scriptsize$A_{3,3}\subset
A_{4,12}$}&{\scriptsize$X_3;X_1,X_2$}&{\scriptsize$X_3=\partial_3+x^1\partial_1 +x^2\partial_2
 $}&{\scriptsize$\eta^{123}=q_1(e^{2x^3}-1)$}\\

{\scriptsize$$}&{\scriptsize$$}&{\scriptsize$X_1=\partial_1,\quad
 X_2=\partial_2$}&{\scriptsize$$}\\

{\scriptsize$$}&{\scriptsize$$}&{\scriptsize$$}&{\scriptsize$$}\\

{\scriptsize$A_{3,6}\subset
A_{4,12}$}&{\scriptsize$X_4;X_1,X_2$}&{\scriptsize$X_1=\partial_4+x^2\partial_1-x^1\partial_2$}&{\scriptsize$\eta^{124}=q_1x^4$}\\

{\scriptsize$$}&{\scriptsize$$}&{\scriptsize$X_1=\partial_1,\quad
 X_2=\partial_2$}&{\scriptsize$$}\\

{\scriptsize$$}&{\scriptsize$$}&{\scriptsize$$}&{\scriptsize$$}\\

{\scriptsize$A^{|a|}_{3,7}\subset
A_{4,12}$}&{\scriptsize$X_4+aX_3;X_1,X_2$}&{\scriptsize$X_4+aX_3=\partial_4+a\partial_3$}&{\scriptsize$\eta^{124}=q_1(e^{x^3+ax^4}-1)$}\\

{\scriptsize$(a\neq0)$}&{\scriptsize$$}&{\scriptsize$+(ax^1+x^2)\partial_1+(ax^2-x^1)\partial_2
 $}&{\scriptsize$\eta^{123}=aq_1(e^{x^3+ax^4}-1)$}\\

 {\scriptsize$$}&{\scriptsize$$}&{\scriptsize$X_1=\partial_1,\quad
 X_2=\partial_2$}&{\scriptsize$$}\\
 \hline\hline
\end{tabular}
\normalsize
\section{\bf Superintegrable systems with Nambu structures}
A Hamiltonian system with $n$ degrees of freedom is integrable from the Liouville sense if it has $n$ invariants in involution \cite{fomen} and is superintegrable if it has additional independent invariants up to $2n-1$ \cite{temp}.
Now in this section we will try to construct superintegrable dynamical systems with use of Nambu structures of order four
on four dimensional real Lie groups. Actually here we consider some systems related to the Nambu structure on some
Lie groups as a phase space (i.e. Lie groups which have symplectic structurs \cite{O}).
We perform this work in two parts, in the first part
we consider the superintegrable systems with ${\bf A_{4,8}}$ as symmetry group, then in part two we consider the systems
with ${\bf A_{4,10}}$ as symmetry group. Note that ${\bf A_{4,8}}$ and ${\bf A_{4,10}}$ are only Lie groups which have invertible $ad$ invariant metrics.
In the following we consider the $(x^1,...,x^4)$ as coordinates of the Lie groups.
\subsection{\bf Superintegrable systems with ${\bf A_{4,8}}$ as symmetry group}
\begin{itemize}
\item
{\bf Lie group} ${\bf A^0_{4,9}}$ {\bf as phase space:}\\
The symplectic  structures on ${\bf A^0_{4,9}}$ is as follows \cite{O}
\begin{eqnarray}
\{x^1,x^4\}=\alpha,\;\;\;\;\;\{x^2,x^3\}=-\alpha,
\end{eqnarray}
where $\alpha$ is an arbitrary nonzero real constant. One can find the following Darboux coordinates:
\begin{eqnarray}
y^{1}=-\frac{1}{\alpha}x^{2},\quad y^{2}=x^{1},\quad y^{3}=x^{3},\quad y^{4}=\frac{1}{\alpha}x^{4},\label{dar-A^{0}_{4,9}}
\end{eqnarray}
such that they satisfy in the following standard Poisson brackets:
\footnote{Note that in the further coming examples the Darboux coordinate satisfy the
standard Poisson brackets $(\ref{dar-pos})$.}
\begin{eqnarray}
\{y^{1},y^{3}\}=1,\quad \{y^{2},y^{4}\}=1.\label{dar-pos}
\end{eqnarray}
 Now using the method mentioned in \cite{cur} one can construct a dynamical system by use of the Nambu structure on Lie group ${\bf A^0_{4,9}}$ as phase space. For this purpose consider the dynamical quantities $Q_a$ as functions of ${x^i}$ such that they satisfy in the following relation:
  \begin{eqnarray}
  \{Q_a,Q_b\}=f^c_{ab}Q_c,\label{braQi}
  \end{eqnarray}
  where $f^c_{ab}$ is the structure constant of the symmetry Lie algebra $A_{4,8}$. Now after some calculation one can rewrite the Nambu 4-brackets (weighted by the structure constants) in terms of Poisson bracket as follows \cite{cur}:
  {\small
     \begin{eqnarray}
     g^{ac}g^{bd}f^e_{cd}\{A,Q_a,Q_b,Q_e\}=3g^{ac}g^{bd}f^e_{cd}\eta\{A,Q_a\}\{Q_b,Q_e\}\nonumber
     =3g^{ac}g^{bd}f^e_{cd}f^f_{be}\eta\{A,Q_a\}Q_f,\nonumber
       \end{eqnarray}
       }
       where $g^{ac}$ is inverse of ad invariant non-degenerate metric on the Lie algebra $A_{4,8}$ and $\eta$ is the Nambu structure on Lie group ${\bf A^0_{4,9}}$ in terms of $y^{i}\quad (i=1,2,3,4)$. After some calculations, we find that for the Lie algebra $A_{4,8}$ the $ad$ invariant metric has
       the following form:
       {\small
 \begin{eqnarray}
 g_{ab}=\left(
 \begin{array}{cccccc}
  0&0&0&s\\
 0&0&-s&0\\
 0&-s&0&0\\
 s&0&0&0
 \end{array}
 \right),\label{metricA_{4,8}}
\end{eqnarray}
}
where $s$ is arbitrary nonzero real constant.
 On the otherhand after some calculations one can find that
 {\small
 \begin{eqnarray}
 g^{ac}g^{bd}f^e_{cd}f^f_{be}Q_aQ_f=\frac{-2}{ s^2}Q_1^2,\label{casimir}
  \end{eqnarray}
  }
  where $Q_1$ is the casimir of $A_{4,8}$\cite{pat}; on the other hand for the dynamical system with symmetry
  Lie algebra $A_{4,8}$, it is proportional to the Hamiltonian of the system. In this respect we have:
  {\small
  \begin{eqnarray}
   g^{ac}g^{bd}f^e_{cd}\{A,Q_a,Q_b,Q_e\}=\frac{-3\eta}{s^2}\{A,H\}=\frac{-3\eta}{s^2}\frac{\partial A}{\partial t},
  \end{eqnarray}
  }
  where $\eta$ is the value of the Nambu structure of the Lie group ${\bf A^{0}_{4,9}}$ and $H=Q_1^2$.
  Therefore the evalution of the dynamical system can be discribed in terms of Nambu structure as follows:
  {\small
  \begin{eqnarray}
  \frac{\partial A}{\partial t}=\frac{-s^{2}}{3\eta}g^{ac}g^{bd}f^e_{cd}\{A,Q_a,Q_b,Q_e\}.\label{dysA4,8}
  \end{eqnarray}
  }
 Now by the use of the following realization of the Lie algebra $A_{4,8}$ \cite{pop} in $\Re^4$
 {\small
\begin{eqnarray}
X_1=\partial_1,\;\;\;\;\;X_2=\partial_2,\;\;\;\;\;X_3=y^2\partial_1,\;\;\;\;\;X_4=y^2\partial_2,\nonumber
 \end{eqnarray}
 }
 where $\partial_i=\frac{\partial}{\partial y^i}$,
 we have the following forms\footnote{Note that we use the quantum mechanical realization for $p_i=-\partial_i$
 , where for our standard Poisson bracket $(\ref{dar-pos})$ we have $p_1=y^3$ , $p_2=y^4$.} for the $Q_i$:
 {\small
 \begin{eqnarray}
 Q_1=-p_1=-y^3,\;\;\;\;\;Q_2=-p_2=-y^4,\;\;\;\;\;Q_3=-y^2p_1=-y^2y^3,\;\;\;\;\;Q_4=-y^2p_2=-y^2y^4,\nonumber
  \end{eqnarray}
  }
  such that they satisfy  $(\ref{braQi})$ by using $(\ref{dar-pos})$.
  The corresponding Hamiltonian can be written  as follows:
  {\small
  \begin{eqnarray}
  H=p_1^2=(y^3)^2.\label{HamilA_{4,8}}
   \end{eqnarray}
   }
   In this way one can describe the dynamics of superintegrable system (with involutive functions e.g. $(H,Q_1,Q_2)$\footnote{In the following examples the
   functions which are involutive can be considered as one of the sets $(H,Q_1,Q_2)$ , $(H,Q_1,Q_3)$ or $(H,Q_1,Q_4)$.}) in terms of Nambu structure as follows\footnote{Note that all of systems obtained in the following have same dynamical
   evaluation as $(\ref{dysA4,8})$ and the same Hamiltonian $(\ref{HamilA_{4,8}})$ (in the Darboux coordinate)
   with differences in the value $\eta$ (the cofficient function of the Nambu structure).}:
   {\small
   \begin{eqnarray}
  \frac{\partial A}{\partial t}=\frac{-s^{2}}{3q_4(e^{-2\alpha y^{4}}-1)}g^{ac}g^{bd}f^e_{cd}\{A,Q_a,Q_b,Q_e\},
  \end{eqnarray}
  }
  where we have used $\eta=q_4(e^{-2x^4}-1)$ (Table I) and Darboux coordinate $(\ref{dar-A^{0}_{4,9}})$.
\item
{\bf Lie group} ${\bf A^{-1}_{4,2}}$ {\bf as phase space:}\\
For this Lie group the symplectic  structure has the following form \cite{O}
{\small
\begin{eqnarray}
\{x^1,x^2\}=2\alpha,\quad\{x^1,x^3\}=-\alpha,\quad\{x^2,x^4\}=\beta e^{-x^4},
\end{eqnarray}
}
where $\alpha$ and $\beta$ are arbitrary nonzero real numbers.
The Darboux coordinate for ${\bf A^{-1}_{4,2}}$  can be found as follows\cite{Abed}
{\small
\begin{eqnarray}
&&y^1=-\frac{e^{x^4}}{\beta}+x^3,\quad y^2=\frac{-2\alpha e^{x^4}-\beta x^1+\alpha\beta x^2}{\alpha\beta^2}\nonumber\\
&&y^3=\frac{2e^{x^4}}{\beta}+\frac{x^1}{\alpha},\quad y^4=e^{x^4},
\end{eqnarray}
}
then the Nambu structure on the Lie group ${\bf A^{-1}_{4,2}}$ (Tabel I) in terms of $y^{i}\quad (i=1,2,3,4)$ is obtained as
$\eta=q_{4}(\frac{1}{y^4}-1)$ so  that from $(\ref{dysA4,8})$ the dynamics of the superintegrable system can be described as follows:
{\small
 \begin{eqnarray}
  \frac{\partial A}{\partial t}=\frac{-s^{2}}{3q_{4}(\frac{1}{y^4}-1)}g^{ac}g^{bd}f^e_{cd}\{A,Q_a,Q_b,Q_e\},\nonumber
  \end{eqnarray}
  }
  with the Hamiltonian:
  {\small
  \begin{eqnarray}
  H=(y^3)^2=(\frac{2e^{x^4}}{\beta}+\frac{x^1}{\alpha})^2.\nonumber
  \end{eqnarray}
  }
  \item
  {\bf Lie group} ${\bf A_{4,3}}$ {\bf as phase space:}\\
  The non-degenarete Poisson structure on ${\bf A_{4,3}}$ is as follows \cite{O}
  {\small
\begin{eqnarray}
\{x^1,x^2\}=\alpha x^4 e^{-x^4},\quad\{x^1,x^3\}=\beta e^{-x^4},\quad\{x^1,x^4\}=\gamma e^{-x^4},\quad\{x^2,x^3\}=\lambda,
\end{eqnarray}
}
where $\alpha$, $\beta$, $\gamma$ and $\lambda$ are arbitrary nonzero real numbers.
The Darboux coordinate for ${\bf A_{4,3}}$ can be obtained as follows\cite{Abed}:
{\footnotesize
\begin{eqnarray}
&& y^1=\frac{\beta x^2}{\lambda}+\frac{\alpha\gamma(x^3)^{2}}{2\beta\lambda}-\frac{\alpha x^3x^4}{\lambda},\quad y^2=\frac{x^1}{\gamma}-\frac{\beta e^{-x^4}x^2}{\lambda\gamma}
-\frac{\alpha e^{-x^4}(x^3)^{2}}{2\beta\lambda}+
\frac{\alpha e^{-x^4}x^3x^4}{\lambda\gamma}\nonumber\\
&& y^3=\frac{x^3}{\beta},\quad y^4=e^{x^4},
\end{eqnarray}
}
in this way the Nambu structure on Lie group ${\bf A_{4,3}}$ (Tabel I) in terms of $y^{i}\quad (i=1,2,3,4)$ is obtained as follows
{\small
\begin{eqnarray}
\eta=q_1\gamma y^2+q_1\frac{y^1}{y^4}+\frac{\alpha q_1y^3\ln y^4}{2\lambda y^4}-\frac{q_1\alpha\beta y^3\ln y^4}{\lambda y^4}+\frac{q_4}{y^4}-q_4,\nonumber
\end{eqnarray}
}
and from $(\ref{dysA4,8})$, the dynamics of the superintegrable system can be described in term of Nambu bracket as follows:
{\footnotesize
\begin{eqnarray}
  \frac{\partial A}{\partial t}=\frac{-s^{2}}{3(q_1\gamma y^2+q_1\frac{y^1}{y^4}+\frac{\alpha q_1y^3\ln y^4}{2\lambda y^4}-\frac{q_1\alpha\beta y^3\ln y^4}{\lambda y^4}+\frac{q_4}{y^4}-q_4)}g^{ac}g^{bd}f^e_{cd}\{A,Q_a,Q_b,Q_e\},
  \end{eqnarray}
  }
  with the Hamiltonian
  {\small
  \begin{eqnarray}
  H=(y^3)^2=(\frac{x^3}{\beta})^2.\nonumber
  \end{eqnarray}
  }
  \item
  {\bf Lie group} ${\bf A^{a,0}_{4,6}}$ {\bf as phase space:}\\
  For this Lie group the non-degenarete Poisson structure has the following form \cite{O}
  {\small
\begin{eqnarray}
\{x^1,x^4\}=\gamma e^{{-\alpha}x^4},\quad\{x^2,x^3\}=\beta,
\end{eqnarray}
}
where $\alpha$, $\beta$ and $\gamma$ are arbitrary nonzero real numbers.
The Darboux coordinate for Lie group ${\bf A^{a,0}_{4,6}}$  has the following form \cite{Abed}
{\small
\begin{eqnarray}
&&y^1=x^3,\quad y^2=\frac{e^{2\alpha x^4}x^1}{\alpha\gamma},
\nonumber\\
&&y^3=-\frac{x^2}{\beta},\quad y^4=e^{-\alpha x^4},
\end{eqnarray}
}
then the Nambu structure on Lie group ${\bf  A^{a,0}_{4,6}}$ (Table I) in terms of $y^{i}\quad (i=1,2,3,4)$ is obtained as follows
{\small
\begin{eqnarray}
\eta=q_1\alpha\gamma(y^4)^2y^2+q_4y^4-q_4,\nonumber
\end{eqnarray}
}
such that from $(\ref{dysA4,8})$ the dynamics of the superintegrable system can be described as follows:
{\small
\begin{eqnarray}
  \frac{\partial A}{\partial t}=\frac{-s^{2}}{3(q_1\alpha\gamma(y^4)^2y^2+q_4y^4-q_4)}g^{ac}g^{bd}f^e_{cd}\{A,Q_a,Q_b,Q_e\},
  \end{eqnarray}
  }
  with the Hamiltonian:
  {\small
  \begin{eqnarray}
  H=(y^3)^2=(\frac{x^2}{\beta})^2.\nonumber
  \end{eqnarray}
  }
  \item
  {\bf Lie group} ${\bf A_{4,7}}$ {\bf as phase space:}\\
  The non-degenarete Poisson structure on ${\bf A_{4,7}}$ is as follows\cite{O}
  {\small
\begin{eqnarray}
\{x^1,x^3\}=-2\alpha x^3e^{{-2}x^4},\quad\{x^1,x^4\}=\alpha e^{{-2}x^4},\quad\{x^2,x^3\}=2\alpha e^{{-2}x^4},
\end{eqnarray}
}
where $\alpha$ is an arbitrary nonzero real number.
The Darboux coordinate for ${\bf A_{4,7}}$  can be found as follows\cite{Abed}
{\small
\begin{eqnarray}
&&y^1=\frac{e^{2x^4}x^2}{2\alpha},\quad y^2=-\frac{-1-e^{2x^4}+e^{4x^4}x^1+e^{4x^4}x^2x^3}{2\alpha},
\nonumber\\
&&y^3=x^3,\quad y^4=e^{-2x^4},
\end{eqnarray}
}
and consequently the Nambu structure on Lie group ${\bf  A_{4,7}}$ (Table I) in terms of $y^{i}\quad (i=1,2,3,4)$ can be obtained as follows:
{\small
\begin{eqnarray}
\eta=q_4((y^4)^2-1),\nonumber
\end{eqnarray}
}
in this way from $(\ref{dysA4,8})$ the dynamical equation of the superintegrable system can be written as follows:
{\small
\begin{eqnarray}
  \frac{\partial A}{\partial t}=\frac{-s^{2}}{3q_4((y^4)^2-1)}g^{ac}g^{bd}f^e_{cd}\{A,Q_a,Q_b,Q_e\},
  \end{eqnarray}
  }
  with the Hamiltonian:
  {\small
  \begin{eqnarray}
  H=(y^3)^2=(x^3)^2.\nonumber
  \end{eqnarray}
  }
\item
  {\bf Lie group} ${\bf A^{1}_{4,9}}$ {\bf as phase space:}\\
The symplectic  structures on ${\bf A^{1}_{4,9}}$ is as follows \cite{O}
{\small
\begin{eqnarray}
\{x^1,x^3\}=2\alpha x^3e^{-2x^4},\quad\{x^1,x^4\}=-\alpha e^{-2x^4},\quad\{x^2,x^3\}=-2\alpha e^{-2x^4},
\end{eqnarray}
}
where $\alpha$ ia an arbitrary nonzero real number.
The Darboux coordinate for the Lie group ${\bf A^{1}_{4,9}}$  have the following forms \cite{Abed}:
{\small
\begin{eqnarray}
&&y^1=-\frac{e^{2x^4}x^2}{2\alpha},\quad y^2=\frac{-1- e^{2x^4}+e^{4x^4}x^1+e^{2x^4}x^2x^3}{2\alpha}\nonumber\\
&&y^3=x^3,\quad y^4=e^{-2x^4},
\end{eqnarray}
}
in this way the Nambu structure on Lie group ${\bf A^{1}_{4,9}}$ (Table I) in terms of $y^{i}\quad (i=1,2,3,4)$ can be written as $\eta=q_{4}((y^4)^2-1)$
 and the dynamical equation of the superintegrable system can be describe as follows:
 {\small
 \begin{eqnarray}
  \frac{\partial A}{\partial t}=\frac{-s^{2}}{3q_{4}((y^4)^2-1)}g^{ac}g^{bd}f^e_{cd}\{A,Q_a,Q_b,Q_e\},\nonumber
  \end{eqnarray}
  }
  with the Hamiltonian:
  {\small
  \begin{eqnarray}
  H=(y^3)^2=(x^3)^2.\nonumber
  \end{eqnarray}
  }
  \item
  {\bf Lie group} ${\bf A_{4,12}}$ {\bf as phase space:}\\
  The non-degenarete Poisson structure on ${\bf A _{4,12}}$ is as follows\cite{O}
  {\footnotesize
\begin{eqnarray}
&&\{x^1,x^3\}=-\gamma e^{-x^3}(\alpha\cos x^4+\beta\sin x^4),\quad\{x^1,x^4\}=\gamma e^{-x^3}(-\beta\cos x^4+\alpha\sin x^4),\nonumber\\
&&\{x^2,x^3\}=\gamma e^{-x^3}(-\beta\cos x^4-\alpha\sin x^4),\;\{x^2,x^4\}=-\gamma e^{-x^3}(\alpha\cos x^4+\beta\sin x^4), \gamma=\frac{1}{\alpha^2+\beta^2}
\end{eqnarray}
}
where $\alpha$ and $\beta$ are arbitrary nonzero real numbers.
The Darboux coordinate for ${\bf A_{4,12}}$ can be found as follows\cite{Abed}
{\small
\begin{eqnarray}
&&y^1=e^{2x^3}(\alpha x^1\cos x^4-\beta x^2\cos x^4+\beta x^1\sin x^4+\alpha x^2\sin x^4),\nonumber\\
&&y^2=e^{-x^3}(\beta x^1\cos x^4+\alpha x^2\cos x^4-\alpha x^1\sin x^4+\beta x^2\sin x^4),\nonumber\\
&&y^3=e^{x^3},\quad y^4=x^4,
\end{eqnarray}
}
then the Nambu structure on Lie group ${\bf  A_{4,12}}$ (Table I) in terms of $y^{i}\quad (i=1,2,3,4)$ is obtained as follows
{\small
\begin{eqnarray}
\eta=q_4(\frac{1}{(y^3)^2}-1),\nonumber
\end{eqnarray}
}
therefore  from $(\ref{dysA4,8})$ the dynamical equation of the superintegrable system can be written as follows:
{\small
\begin{eqnarray}
  \frac{\partial A}{\partial t}=\frac{-s^{2}}{3q_4(\frac{1}{(y^3)^2}-1)}g^{ac}g^{bd}f^e_{cd}\{A,Q_a,Q_b,Q_e\},
  \end{eqnarray}
  }
  with the Hamiltonian:
  {\small
  \begin{eqnarray}
  H=(y^3)^2=e^{2x^3}.\nonumber
  \end{eqnarray}
  }
  \end{itemize}
  \subsection{\bf Superintegrable systems with ${\bf A_{4,10}}$ as a symmetry group}
  As in the previous section we construct some dynamical systems by use of the Nambu structure on
  some Lie groups as a phase space and with symmetry group ${\bf A_{4,10}}$. One can find the $ad$ invariant non-degenerate
  metric on the Lie algebra $A_{4,10}$ as follows:
  {\small
  \begin{eqnarray}
 g_{ab}=\left(
 \begin{array}{cccccc}
  0&0&0&m\\
 0&m&0&0\\
 0&0&m&0\\
 n&0&0&0
 \end{array}
 \right),\label{metricA_{4,10}}
\end{eqnarray}
}
where $m$ and $n$ are arbitrary nonzero real numbers.\\
For this case we have the following relations insted of $(\ref{casimir})$ and $(\ref{dysA4,8})$:
{\small
 \begin{eqnarray}
 g^{ac}g^{bd}f^e_{cd}f^f_{be}Q_aQ_f=\frac{2}{mn}Q_1^2,
  \end{eqnarray}
  }
  {\small
   \begin{eqnarray}
  \frac{\partial A}{\partial t}=\frac{mn}{3\eta}g^{ac}g^{bd}f^e_{cd}\{A,Q_a,Q_b,Q_e\},\label{dysA4,10}
  \end{eqnarray}
  }
  such that by use of the following realization \cite{pop} in $\Re^4$:
  {\small
  \begin{eqnarray}
  X_1=\partial_1,\;\;\;X_2=\partial_2,\;\;\;X_3=y^1\partial_1+y^2\partial_2,\;\;\;X_4=y^2\partial_1-y_1\partial_2,
  \end{eqnarray}
  }
  we have
  {\small
  \begin{eqnarray}
  Q_1=-y^3,\;\;\;Q_2=-y^4,\;\;\;Q_3=-y^1y^3-y^2y^4,\;\;\; Q_4=-y^2y^3+y^1y^4,
  \end{eqnarray}
  }
  then the Hamiltonian of the superintegrable systems with symmetry group ${\bf A_{4,10}}$ are the same as in the
  previous section with the differences in the dynamical equations $(\ref{dysA4,8})$ (i.e. in the inverse metrics of $(\ref{metricA_{4,8}})$
  and $(\ref{metricA_{4,10}})$ in $(\ref{dysA4,8})$ and also structure constants $f^{c}_{ab}$ of $A_{4,8}$ and $A_{4,10}$).\\
  In this way the dynamical equations for the real four dimensional Lie groups as phase space are written as follows:
  \begin{itemize}
  \item ${\bf A^0_{4,9}}$ {\bf as phase space}
  {\footnotesize
  \begin{eqnarray}
  \frac{\partial A}{\partial t}=\frac{mn}{3q_4(e^{-2\alpha y^{4}}-1)}g^{ac}g^{bd}f^e_{cd}\{A,Q_a,Q_b,Q_e\},
  \end{eqnarray}
  }
  \item ${\bf A^{-1}_{4,2}}$ {\bf as phase space}
  {\footnotesize
  \begin{eqnarray}
  \frac{\partial A}{\partial t}=\frac{mn}{3q_{4}(\frac{1}{y^4}-1)}g^{ac}g^{bd}f^e_{cd}\{A,Q_a,Q_b,Q_e\},
  \end{eqnarray}
  }
  \item ${\bf A_{4,3}}$ {\bf as phase space}
  {\footnotesize
  \begin{eqnarray}
  \frac{\partial A}{\partial t}=\frac{mn}{3(q_1\gamma y^2+q_1\frac{y^1}{y^4}+\frac{\alpha q_1y^3\ln y^4}{2\lambda y^4}-\frac{q_1\alpha \beta y^3\ln y^4}{\lambda y^4}+\frac{q_4}{y^4}-q_4)}g^{ac}g^{bd}f^e_{cd}\{A,Q_a,Q_b,Q_e\},
  \end{eqnarray}
  }
  \item ${\bf A^{a,0}_{4,6}}$ {\bf as phase space}
  {\footnotesize
  \begin{eqnarray}
  \frac{\partial A}{\partial t}=\frac{mn}{3(q_1\alpha\gamma(y^4)^2y^2+q_4y^4-q_4)}g^{ac}g^{bd}f^e_{cd}\{A,Q_a,Q_b,Q_e\},
  \end{eqnarray}
  }
  \item ${\bf A_{4,7}}$ {\bf as phase space}
  {\footnotesize
  \begin{eqnarray}
 \frac{\partial A}{\partial t}=\frac{mn}{3q_4((y^4)^2-1)}g^{ac}g^{bd}f^e_{cd}\{A,Q_a,Q_b,Q_e\},
 \end{eqnarray}
 }
 \item ${\bf A^{1}_{4,9}}$ {\bf as phase space}
 {\footnotesize
 \begin{eqnarray}
  \frac{\partial A}{\partial t}=\frac{mn}{3q_{4}((y^4)^2-1)}g^{ac}g^{bd}f^e_{cd}\{A,Q_a,Q_b,Q_e\},
  \end{eqnarray}
  }
  \item ${\bf A_{4,12}}$ {\bf as phase space}
  {\footnotesize
  \begin{eqnarray}
  \frac{\partial A}{\partial t}=\frac{mn}{3q_4(\frac{1}{(y^3)^2}-1)}g^{ac}g^{bd}f^e_{cd}\{A,Q_a,Q_b,Q_e\}.
  \end{eqnarray}
  }
  \end{itemize}

\vspace{.7cm}
\section*{\bf Conclusions}
We have obtained all Nambu structures of order four and three on four dimensional real Lie groups.
Also we have obtained new superintegrable systems with the some four dimensional real Lie groups as
phase space and symmetry group ${\bf A_{4,8}}$ and ${\bf A_{4,10}}$ such that their dynamical evaluation equation
are described in terms of related Nambu structures of order four.


 \vspace{.7cm}
\subsection*{Acknowledgment}
\vspace{3mm} We would like to express our heartiest  gratitude to V. Barzgar and F. Darabi for carefully reading
the manuscript and their useful comments.

\end{document}